\newcounter{myctr}
\def\myitem{\refstepcounter{myctr}\bibfont\noindent\ifnum\themyctr>9\else\phantom{0}\fi\hangindent17pt\themyctr.\enskip}

\documentclass{ws-ijqi}
\usepackage{slashbox}

\begin{document}

\markboth{Haoyang Wu} {On the justification of applying quantum
strategies to the Prisoners' Dilemma and mechanism design}

%%%%%%%%%%%%%%%%%%%%% Publisher's Area please ignore %%%%%%%%%%%%%%
\catchline{}{}{}{}{}
%%%%%%%%%%%%%%%%%%%%%%%%%%%%%%%%%%%%%%%%%%%%%%%%%%%%%%%%%%%%%%%%%%%

\title{On the justification of applying quantum strategies to the Prisoners' Dilemma and mechanism design}

\author{Haoyang Wu}

\address{Department of Physics, Xi'an Jiaotong University,\\
Xi'an, 710049, China.\\
hywch@mail.xjtu.edu.cn}

\maketitle

\begin{history}
%\received{13 Apr 2010}
%\revised{ 2010}
%\accepted{ 2010}
%\comby{(xxxxxxxxxx)}
\end{history}

\begin{abstract}
The Prisoners' Dilemma is perhaps the most famous model in the field
of game theory. Consequently, it is natural to investigate its
quantum version when one considers to apply quantum strategies to
game theory. There are two main results in this paper: 1) The
well-known Prisoners' Dilemma can be categorized into three types
and only the third type is adaptable for quantum strategies. 2) As a
reverse problem of game theory, mechanism design provides a better
circumstance for quantum strategies than game theory does.
\end{abstract}


\section{Introduction}

Game theory and mechanism design are two important branches of
economics. Game theory aims to investigate rational decision making
in conflict situations, whereas mechanism design just concerns the
\emph{reverse} question: given some desirable outcomes, can we
design a game that produces it?

In 1999, Eisert \emph{et al} $^{1}$ proposed a pioneering
quantum-version model of two-player Prisoners' Dilemma. The novel
model showed a fascinating ``quantum advantages'' as a result of a
novel quantum Nash equilibrium. Guo \emph{et al} $^{2}$ gave a good
review on quantum games. In 2010, Wu $^{3}$ investigated what would
happen if agents could use quantum strategies in the theory of
mechanism design. The result was interesting, i.e., by virtue of a
quantum mechanism, agents who satisfied a certain condition could
combat ``bad'' social choice rules instead of being restricted by
the traditional mechanism design theory.

Despite the aforementioned accomplishments in quantum games, in
2002, van Enk and Pike $^{4}$ criticized that a quantum game was
indeed a new game that was constructed and solved, not the original
classical game. In this paper, we will deeply investigate whether
and when quantum strategies are useful for economic society. In
Section 2, we will analyze the justification of applying quantum
strategies to the Prisoners' Dilemma. In Section 3, we will
investigate the justification of applying quantum strategies to
mechanism design. Section 4 draws the conclusions.

\section{The justification of applying quantum strategies to the Prisoners' Dilemma}
\subsection{Three types of Prisoners' Dilemma}
As is well known, the Prisoners' Dilemma is a simple model that
captures the essential contradiction between individual rationality
and global rationality. In the classical Prisoners' Dilemma, two
prisoners are arrested by a policeman. Each prisoner must
independently choose to cooperate (strategy $C$) or to defect
(strategy $D$). Table 1 shows the payoff matrix of two prisoners,
Alice and Bob.

\emph{Table 1: The payoff matrix of two prisoners. The first entry in the
parenthesis denotes the payoff of Alice and the second
stands for the payoff of Bob}.\\
\begin{tabular}{|c|c|c|}
\hline \backslashbox{$Alice$}{$Bob$} & {$C$}&{$D$}
 \\\hline $C$ & (3, 3) & (0, 5)
\\ $D$ & (5, 0) & (1, 1)
\\ \hline
\end{tabular}

Suppose the prisoners are rational, then the unique Nash equilibrium
is the dominant strategy ($D$, $D$), while the Pareto optimal
strategy is ($C$, $C$). Although the Prisoners' Dilemma has found
widespread applications in many disciplines such as economics,
politics, sociology and so on, perhaps not everybody notice that
actually there are three different types of Prisoners' Dilemma:

\textbf{Type-1 Prisoners' Dilemma: } There is no arbitrator to
assign payoffs to the players, i.e., there are only two players in
the game, whose payoffs are \emph{generated} by the players
themselves. For example, let us consider two countries (e.g., US and
Russia) confronted the problem of nuclear disarmament. The strategy
$C$ means ``Obeying disarmament'', and $D$ means ``Refusing
disarmament''. Obviously, in the nuclear disarmament game, there is
no arbitrator and the payoffs of two countries are generated by
themselves.

\textbf{Type-2 Prisoners' Dilemma: } There is an arbitrator in the
game to assign payoffs to the players. Before the players send
strategies to the arbitrator independently, \emph{they cannot
communicate to each other}. For example, if the prisoners are
arrested separately, this is the type-2 Prisoners' Dilemma.

\textbf{Type-3 Prisoners' Dilemma: } There is an arbitrator in the
game to assign payoffs to the players. Before the players send
strategies to the arbitrator independently, \emph{they can
communicate to each other but cannot enter into a binding contract}.
For example, if the prisoners are arrested in one room and confront
a payoff matrix specified by Table 1, this is the type-3 Prisoners'
Dilemma. It should be emphasized that although the rational players
all agree that the strategy $C$ will benefit both of them, they will
definitely choose the strategy $D$ when they actually send
strategies to the policeman independently, because ($D$, $D$) is the
unique dominant strategy. This is the catch of the dilemma. (Note:
If the two players can enter into a binding contract before they
send strategies to the arbitrator, then the game will become a
cooperative game. We simply ignore this case.)

\subsection{Eisert \emph{et al}'s model}

In 1999, the Prisoners' Dilemma was generalized to a quantum domain
$^{1}$. Fig. 1 shows the setup of a two-player quantum game. The
timing sequence of the quantum Prisoners' Dilemma is referred to
Ref. 1. Under classical circumstances, people usually do not
discriminate the differences among three types of Prisoners' Dilemma
because they correspond to the same payoff matrix (i.e., Table 1).
But in the context of quantum domain, only the type-3 Prisoners'
Dilemma is adaptable for quantum strategies. In the following, we
explain the three cases in details.

\begin{figure}
\centering
\includegraphics[height=1.4in,clip,keepaspectratio]{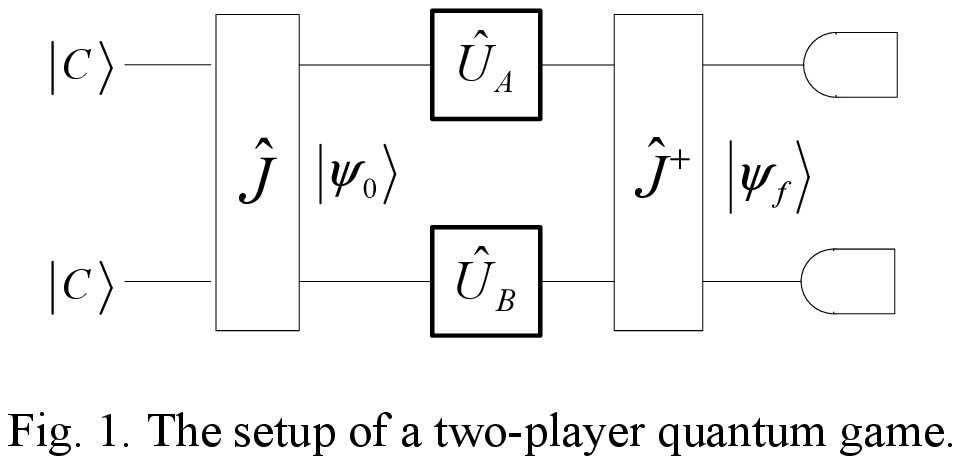}
%\caption{IT2 fuzzy sets corresponding to the information system
%given in Table 1}
%\label{fig:graph}
\end{figure}

\textbf{Type-1 Prisoners' Dilemma: } Obviously, it is meaningless to
consider the quantum version of type-1 Prisoners' Dilemma, because
in Eisert \emph{et al}'s model, the payoffs of two players are
assigned by an arbitrator, which is inconsistent to the model of
type-1 Prisoners' Dilemma.

\textbf{Type-2 Prisoners' Dilemma: } In Eisert \emph{et al}'s
original paper, it is implicit whether the quantum version of the
type-2 Prisoners' Dilemma is feasible or not. Here, we will analyze
this problem deeply. Initially, the two qubits possessed by the
prisoners are separable. It is impossible for the prisoners
themselves to entangle the two qubits, because they cannot
communicate to each other and each prisoner can only perform a local
unitary operation on his/her own qubit. Therefore, we have to assume
that the policeman entangles the two qubits. However, just as van
Enk and Pike have said $^{4}$: ``.. it seems to us counter to the
spirit of the game to have an attorney or interrogater be helpful to
the prisoners and give them an entangled state'', it is unreasonable
to make this assumption.

Furthermore, there are two ways to understand the unreasonableness:
1) Since the Prisoners' Dilemma is a game with complete information,
the policeman knows exactly that once the prisoners possess two
entangled qubits, they will certainly choose the quantum strategy
($\hat{Q}, \hat{Q}$) and obtain the Pareto optimal payoffs (3,3).
Therefore, if the policeman does help the prisoners perform
entangling operation, then \emph{why doesn't the policeman assign
the Pareto optimal payoffs} (3,3) \emph{directly to the prisoners?}
2) Since the role of policeman is the arbitrator, it is unreasonable
to require the policeman to have an incentive to help the prisoners
better off and reach their Pareto optimal payoffs.

Consequently, the Eisert \emph{et al}'s model is meaningless to the
type-2 Prisoners' Dilemma.

\textbf{Type-3 Prisoners' Dilemma: } For this case, the two
prisoners can communicate to each other. Hence, it is reasonable to
assume that the two qubits hold by the prisoners can be entangled by
themselves, i.e., the quantum game doesn't need a benevolent
attorney or interrogater to give the prisoners an entangled state.
The essence of the quantum game is that for the type-3 Prisoners'
Dilemma, the unique dominant strategy is changed from ($\hat{D}$,
$\hat{D}$) to ($\hat{Q}$, $\hat{Q}$), which results in a Pareto
optimal payoff (3,3).

Now, we have categorized the well-known Prisoners' Dilemma into
three different types. It can be seen that Eisert \emph{et al}'s
model is adaptable only to the type-3 Prisoners' Dilemma. Can we
think that the problem about the justification of applying quantum
strategies to the Prisoners' Dilemma has been solved? Not yet,
because another serious question arises naturally: Since the
policeman also knows the aforementioned three types of Prisoners'
Dilemma, \emph{then why does he allow the prisoners to communicate
to each other and have a chance to use quantum strategies}? For the
case of the classical Prisoners' Dilemma, there is no reason for the
policeman to do so.

\section{The justification of applying quantum strategies to mechanism design}
\subsection{Maskin's classical theorem}
In the field of economics, game theory has a \emph{reverse} problem,
i.e., the theory of mechanism design. Game theory aims to predict
the outcome of a given game, whereas the theory of mechanism design
concerns a ``\emph{social engineering}'' question: given some
desirable outcomes, can we design a game that produces it?

There are two important notions in mechanism design theory: social
choice rule (SCR) and mechanism. An SCR specifies the objects that
the designer would like to implement: in each state, she would like
to realize some set of outcomes, but unfortunately, she does not
know the true state $^{5}$. A mechanism is a representation of the
social institution through which the agents interact with the
designer and with one another: each agent sends a message to the
designer, who chooses an outcome as a function of these strategy
choices $^{6}$.

Ref. 5 is a fundamental work in the field of mechanism design. It
provides an almost complete characterization of social choice rules
(SCRs) that are Nash implementable. The main results of Ref. 5 are
the two following theorems: 1) (\emph{Necessity}): If an SCR $F$ is
Nash implementable, then it is monotonic. 2) (\emph{Sufficiency}):
Let $n\geq 3$, if an SCR $F$ is monotonic and satisfies no-veto,
then it is Nash implementable.

According to the sufficiency theorem, even if all agents dislike an
SCR specified by the designer, as long as it is monotonic and
satisfies no-veto, one can always construct a mechanism to implement
the SCR in Nash equilibrium.

\subsection{Wu's quantum mechanism}
In 2010, Wu investigated what would happen if agents could use
quantum strategies in the theory of mechanism design $^{3}$. The
setup of a quantum mechanism is depicted in Fig. 1. Working steps of
the quantum mechanism are referred to Ref. 3.
\begin{figure}[!t]
\centering
\includegraphics[height=2.5in,clip,keepaspectratio]{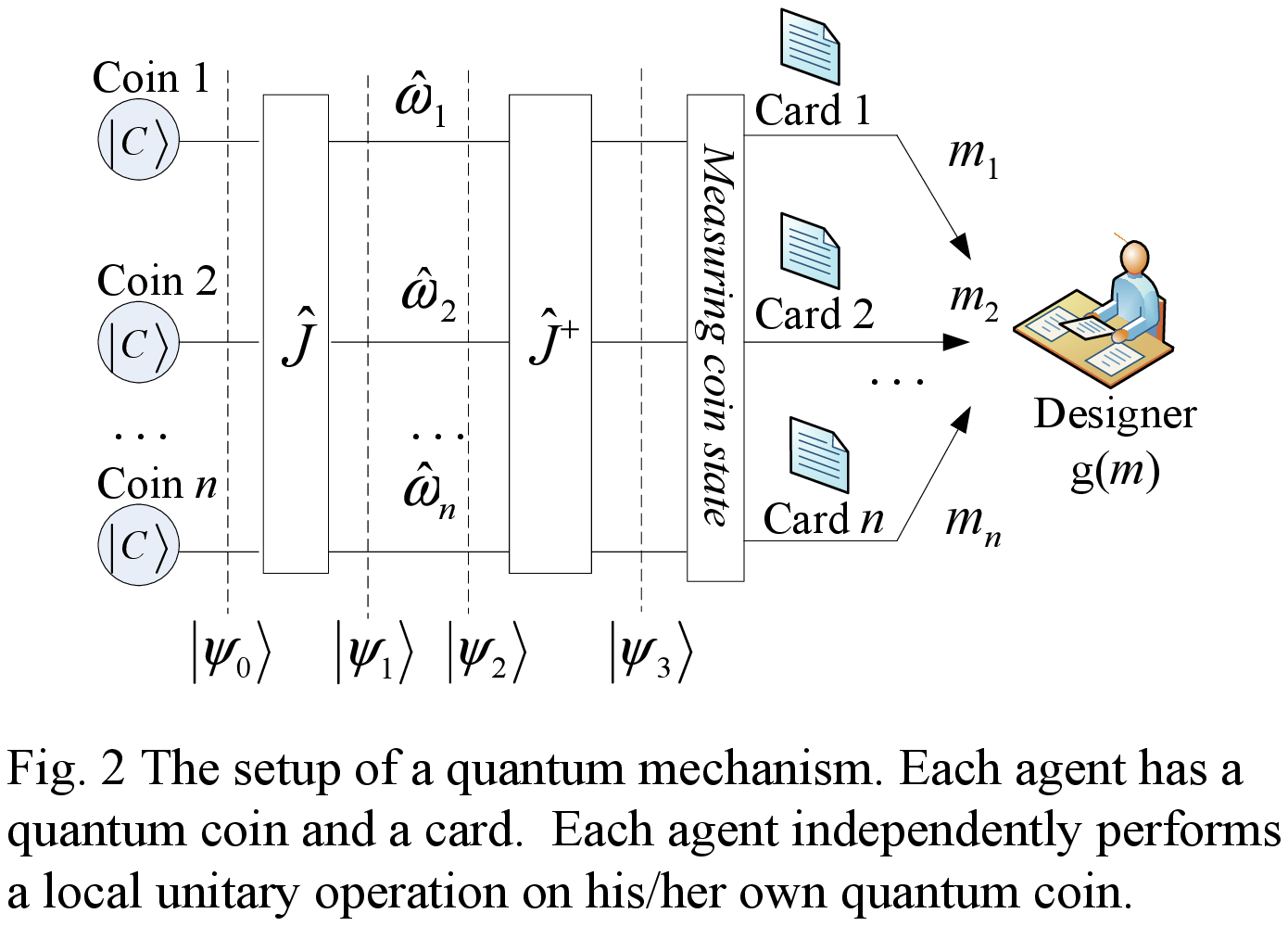}
%\caption{IT2 fuzzy sets corresponding to the information system
%given in Table 1}
%\label{fig:graph}
\end{figure}

According to Ref. 3, when condition $\lambda$ is satisfied, an
original Nash implementable ``bad'' (i.e., Pareto-inefficient) SCR
will no longer be Nash implementable in the context of quantum
domain. The key point of the quantum mechanism is to carry out a
quantum version of \emph{n}-player type-3 Prisoners' Dilemma. In the
end of Section 2, we argue that the type-3 Prisoners' Dilemma is not
meaningful either. However, according to the three reasons listed
below, it is natural for us to adopt the quantum version of the
type-3 Prisoners' Dilemma into the field of mechanism design.

1) In the theory of mechanism design, there always exists a designer
(i.e., the arbitrator) to assign outcomes to the agents. Hence, the
payoffs of the agents are specified by the arbitrator.

2) When Nash implementation is concerned, agents possess complete
information. Therefore, it is reasonable to think that agents can
communicate to each other before they send their strategies to the
designer independently.

3) In the theory of mechanism design, the designer is at an
information disadvantage with respect to the agents. He cannot
restrict agents from communicating to each other, so the question in
the end of Section 2 is solved. In Wu's quantum mechanism, from the
viewpoint of the designer, the interface between agents and the
designer is the same as that in the Maskin's classical mechanism.
The designer cannot discriminate whether the underlying principle of
agent's actions is classical or quantum mechanical. Put in other
words, \emph{the designer cannot restrict the agents from using
quantum strategies to reach their Pareto-optimal outcomes}.

\section{Conclusions}
In this paper, we categorize the well-known Prisoners' Dilemma into
three types, and point out that Eisert \emph{et al}'s model is
adaptable only to the type-3 Prisoners' Dilemma. Moreover, in the
context of game theory, it is still unreasonable that the arbitrator
allow the agents to communicate to each other and have a chance to
use quantum strategies. Just as Maskin $^{7}$ has said: ``Mechanism
design provides the circumstances perhaps most favorable for Nash
equilibrium being a good predictor of human behavior in strategic
settings.'' , in Section 3, we find that mechanism design provides a
more favorable circumstance for quantum strategies than game theory
does.

%--------------------------------------------------------

\end{document}